# Scalar theory of gravity as a pressure force

Mayeul Arminjon

Laboratoire "Sols, Solides, Structures",
Institut de Mécanique de Grenoble, France

**Abstract-** The construction of this theory starts with Newtonian space-time and a tentative interpretation of gravity as Archimedes' thrust exerted on matter at the scale of elementary particles by an imagined perfect fluid or ether. This leads to express the gravity acceleration by a simple formula in which the "ether pressure" $p_e$ plays the role of the Newtonian potential. The instantaneous propagation of Newtonian gravity is obtained with an incompressible ether, giving a field equation for $p_e$. For a compressible ether, this equation holds in the static case and results in a non-linear influence of the mass distribution. The extension of the field equation to non-static situations follows the lines of acoustics and leads to gravitational (pressure) waves. To account for metric effects, first the modern version of the Lorentz-Poincaré interpretation of special relativity is summarized. Then Einstein's equivalence principle (EP) is seen as a correspondence between the metric effects of gravity and those of uniform motion with respect to the ether : a gravitational contraction (dilation) of space (time) standards is assumed, and it implies geodesic motion for test particles in a static field. The same field equation is now expressed in terms of the physical space and time metrics in the frame of ether; since both metrics depend on $p_e$ due to the EP, it becomes non-linear in $p_e$. In the spherical static situation, Schwarzschild's exterior metric is predicted, but the interior metric differs from GR, though it agrees with the EP. Since pressure produces gravitation also in the proposed theory, no equilibrium is possible for very massive objects. But the gravitational collapse in free fall does not lead to any singularity. Moreover, gravitational waves, and even shock ones, can exist *in vacuo* also with spherical symmetry.

## 1. INTRODUCTION

Of the reasons that may justify the search for alternatives to general relativity (GR), some are comparatively recent : the complete implosion into a point singularity, which is predicted for very massive objects and which could not be detected in the case of spherical symmetry; the isolation of gravitation with respect to the theories of other interactions (this expressed earlier in the problem of unifying gravitation with electromagnetism, and more recently in the difficulties with quantum gravity); and, finally, the problems of interpretation which are due to the need for a non-covariant coordinate condition : Papapetrou [1] already suggested that different coordinate conditions might lead to different post-Newtonian equations of motion, and Fock [2] observed (at § 96) that indeed, the coordinates used by Einstein and Infeld differed from harmonic coordinates, used by Fock and Papapetrou, only by second-order $(O(1/c^4))$ terms which play no role in the first corrections to Newton's theory; today the post-Newtonian calculations are very generally formulated in terms of harmonic coordinates [3, 4]. According to Logunov *et al.* [5], one has to modify GR into a "relativistic theory of gravitation" such that, essentially, Fock's harmonic condition is organically contained in the new theory, for otherwise GR would not furnish unique predictions. Thus the general covariance is not a *sine qua non* condition; in this paper, a theory with a privileged frame is investigated. In the writer's opinion, the most fundamental question that remains unsolved by GR was already posed to Newton's theory : how can interactions propagate through empty space ? Why not try to give physical properties to empty space, in such a way that the propagation of gravity as well as the existence of *global* inertial frames (i.e. as they appear in Newtonian theory) become understandable? The existence of the zero-point fluctuations makes it necessary to admit that *vacuum* indeed has physical properties [6,7]. A comparison of Winterberg's ether theory of gravitation [7] with the one proposed here, which will be designated by ETG for short, is deferred to the Conclusion for convenience.

Already within the Newtonian concept of absolute space and time, gravity can be tentatively interpreted as Archimedes's thrust, due to the pressure of an hypothetical, perfect fluid or ether, surrounding all elementary particles of matter : the condition of validity is that all particles should have the same mass density - or at least that their average density depends only on the pressure in the fluid [8a]. It turns out that Newtonian gravity (NG) was first interpreted in this way a long time ago by Euler [9]. Under the condition that NG must be recovered for an incompressible fluid, a (scalar) field equation is obtained for the ether pressure; after a compressibility is introduced, this "ETG" gives qualitatively correct modifications to NG : a finite wave speed and an advance in the perihelion of planetary orbits which is proportional to the prediction of GR [8a]. To go further, it is necessary to have a description of the relativistic effects. The first step is to realize that the old Lorentz-Poincaré ether interpretation of special relativity can be updated with the same logical thrift and consistency that have made Einstein's theory so convincing [10,11]. The "new" ether interpretation reconciles Lorentz's original view of a true contraction in the absolute motion with respect to the ether, with the reciprocity of the Lorentz transformation in the version found by Einstein; it will be once more summarized in this paper. The possibility of this interpretation and its great interest were recognized by Einstein [12] himself (see Builder [10a]). Hereafter, the transition is made to the gravitational situation in which the ether pressure (or equivalently the ether density, since the ether postulated here is a barotropic perfect fluid) varies from one point to another - which is the cause of gravity in the proposed ETG. Special relativity holds then only locally, i.e. in a domain where the ether density may be considered uniform. The true Lorentz space-contraction and time-dilation with motion lead naturally to a new formulation of Einstein's equivalence principle, within the ETG : gravitational space-contraction and time-dilation are postulated; with this assumption and when the proposed equation for the field of ether pressure is reinterpreted as relative to the physical, distorted space and time standards, Schwarzschild's exterior space-time metric is exactly recovered in the static situation with spherical symmetry [8b]. Moreover, for any static gravitation field, the geodesic characterization of the motion of free mass points follows from this assumption and Newton's second law (with the momentum including the velocity-dependent mass and all being measured with the local clock and measuring rod), independently of the field equation.

In this paper, a synthetic presentation of this scalar theory of gravitation is given. Several new results are reported : a discussion of the Newtonian limit; an extension to light-like particles of the proof that Newton's second law (interpreted as above) implies geodesic trajectories; a brief study of gravitational collapse and gravitational waves with spherical symmetry. For these two last problems, the situation in this theory differs strongly from that in GR.

## 2. THE ETHER THEORY OF GRAVITATION WITHIN NEWTONIAN MECHANICS

Following physicists of the past like Euler, Helmholtz and Kelvin, it is found that only a perfect fluid, i.e. without viscosity, could fill the space left by matter without braking any motion : this is well-known in fluid mechanics as d'Alembert's paradox. Such a fluid acts only by its pressure $p_e$; it would exert over a material domain $\Omega$, the force

$$\mathbf{F} = \int_{\partial\Omega} -p_e \, \mathbf{n} \, dS = - \int_{\Omega} \text{grad} \, p_e \, dV \qquad (1)$$

($\partial\Omega$ is the boundary of $\Omega$, $\mathbf{n}$ is the outward normal, S and V are the surface and volume measures). We want that $\mathbf{F}$ is the gravitation force :

$$\mathbf{F} = \int_{\Omega} \rho \, \mathbf{g} \, dV \quad , \qquad (2)$$

with $\mathbf{g}$ the gravity acceleration and $\rho$ the mass density. If we think of $\Omega$ as being a macroscopic domain where $\mathbf{g}$ may be considered uniform (and for the real gravity this is true even with rather large domains), this equality cannot hold, since $\mathbf{g}$ in (2) and grad $p_e$ in (1) should not depend on the kind and state of matter in $\Omega$, while $\rho$ of course does. However, the macroscopic matter is made of particles which are already subjected to the gravitation, hence the fluid pressure would have to act only on a small part of the macroscopic volume $\Omega$ : the union of the volumes $\omega_i$ (i=1,N($\Omega$)) occupied by the constitutive particles. In other words, the fluid would fill the place left by the particles. Let $\rho_i$ be the mass density of particle (i). In order that (1) and (2) coincide when $\Omega=\omega_i$ and $\rho=\rho_i$, it is necessary that $\rho_i$ be independent of the particle (i), and then the pressure force (1) would also be a mass force (2) for a macroscopic domain $\Omega$, provided that grad $p_e$ also be uniform in $\Omega$ :

$$\mathbf{F} = - \text{grad } p_e \sum V(\omega_i) = \mathbf{g} \sum m_i = \mathbf{g} \sum \rho_i V(\omega_i) . \qquad (3)$$

Thus, we come to the conclusion that one should have :

$$\mathbf{g} = - \text{grad } p_e /\rho_p , \qquad (4)$$

with $\rho_p$ the assumed common density of the particles - which could still depend on the pressure $p_e$ in the thought ether. Now the identity between inertial and passive gravitational mass does not seem to be known with the same precision for elementary particles than for macroscopic matter. The density in the particles might thus be allowed to vary from one particle to another, so we take for $\rho_p$ the *average* mass density in the different particles of a macroscopic domain $\Omega$ : $\rho_p= \sum \rho_i V(\omega_i)/ \sum V(\omega_i)$. But in order that $\mathbf{g}$ in (4) does not depend on the kind of present matter, $\rho_p$ must be a function of the state parameters of the ether alone. Since the ether is assumed to be perfectly continuous at any scale, no temperature and no entropy can be defined for it, so the state parameters reduce to the pressure $p_e$ and the "mass" density $\rho_e$ which are related together by the barotropic state equation [13]; hence $\rho_p$ must depend only on $p_e$. This seems to be possible only if the particles of matter are themselves ether : from the attempt to describe gravity as a pressure force in a fluid ether, we are lead to the assumption of a *constitutive ether*, which was a priori set by Romani [13].

Assuming therefore that $\rho_e = \rho_p$ , we obtain the gravity acceleration as

$$\mathbf{g} = - \frac{1}{\rho_e(p_e)} \text{grad } p_e , \qquad (5)$$

but this equation may be taken as a *phenomenological equation* for $\mathbf{g}$ (alternative to the Newtonian equation $\mathbf{g} = \text{grad } U$ ; note that in Newtonian theory, the potential U, as here the pressure $p_e$ of the imagined fluid, manifests itself only through $\mathbf{g}$). The foregoing microscopic considerations are then a heuristic justification for Eq. (5). We note that $p_e$ and $\rho_e$ should be the *macroscopic* pressure and density in the thought ether, since the gravitation varies only over macroscopic distances and due to the effect of macroscopic bodies. This remark is important in the context of a possible unification of the four known physical interactions within a theory of this barotropic ether : the other interactions, which vary over shorter distances, would then have to be all contained in the microscopic fluctuations of ether pressure and velocity. In fact, Romani [13] described a way in which this may be done, in assuming that all known particles would be vortex tori or complexes of such flows; he proposed to explain the appearance of quantum numbers from the requirement that the helicoidal trajectories on the torus should be closed. Yet his work, while very interesting, suffers in some places from a lack of rigour, and contains no field equation. Furthermore, gravity was the least developed part of his work and if he attributed this force to a gradient of ether density in the barotropic ether [13, vol. 1], he stated that the deviation of light rays passing near the Sun should be due to an increase of this density towards the Sun [13, vol. 2] - whereas it follows immediately from our Eq. (5) that $p_e$ and $\rho_e$ *decrease* towards the gravitational attraction. *In the present work, our aim is to deduce from the assumption of the barotropic constitutive ether a phenomenological theory for gravity, in a classical framework, i.e. without trying to obtain a quantized gravity, and also without any claim of describing the other interactions*. Inasmuch as a theory of gravity has to be compared with NG and GR which are also phenomenological theories, we thus do not care about the fact that, of course, quantum mechanics would scarcely allow to define a mass density for an elementary particle, and would not allow to use classical mechanics at this level : within the Newtonian space-time, the theory describes gravity by a potential (Eq. (17a) below) which might be used as a phenomenological description of this interaction, also at the quantum level, in non-relativistic quantum mechanics. Also, once modified to account for "relativistic" effects, the investigated theory, just like GR, will finally lead to a curved space-time, the curvature being bound to the distribution of matter; this is in turn adapted to describe gravity in relativistic quantum mechanics.

In this phenomenological view, the next step is to recover Newtonian gravity. This is a straightforward task, for NG propagates with infinite velocity and thus must correspond to the incompressible case; with a uniform $\rho_e$, we get from Eq. (5) by using Poisson's equation (div $\mathbf{g}$ = -4 $\pi$ G $\rho$ with G Newton's gravitation constant) :

$$\Delta p_e = 4\pi \, G \, \rho \, \rho_e \quad . \qquad\qquad\qquad (6)$$

With a uniform $\rho_e$, Eqs. (5) and (6) give exactly the same description of the field $\mathbf{g}$ as the usual equations $\Delta U$ = -4 $\pi$ G $\rho$ and $\mathbf{g}$ = grad U. The point is that (5) and (6), together with the barotropic equation $\rho_e = \rho_e(p_e)$ (of which the incompressible case is a degenerate one), make sense as well if the ether is compressible (i.e. when $\rho_e$ is an increasing function of $p_e$). In that case, however, we have a new, non-linear field equation for gravity : the fields $p_e$ and $\mathbf{g}$ depend non-linearly of the matter distribution $\rho$, even if the $p_e$-$\rho_e$ relationship is linear. The static case with spherical symmetry has been investigated analytically : one obtains a complete solution that tends towards the Newtonian one as the compressibility tends towards zero, and the exterior solution gives an advance in the perihelion of elliptic orbits which is proportional to the prediction of GR [8a]; in ref. [8a], an extension of Newtonian mechanics to *fluid* inertial frames is also proposed, consistently with the assumption that the frame bound to the macroscopic motion of the fluid ether (the "macro-ether") defines an inertial frame. In the case where the relationship is assumed to be $p_e = \rho_e c^2$ with c the velocity of light (as we will be lead to, Sect. 4), Eq. (6) writes

$$\Delta p_e = (4\pi \, G/c^2) \, \rho \, p_e \qquad\qquad (6bis)$$

which is formally identical to a relativistic equation proposed for static situations by Einstein [14] before he finally came to GR; yet in Einstein's equation the field was of course not $p_e$ but instead the velocity of light w, as expressed in terms of a "universal" time coordinate [this circumstance was not known to the writer for the work [8]; the writer learned this in the papers by Soos [15], who presents a systematic analysis of Kepler's problem in Einstein's successive theories of gravitation].

A motion of massive bodies causes a disturbance in the gravitation field, i.e. in the field of ether pressure $p_e$; but with a compressible ether, this disturbance in the field $p_e$ cannot propagate instantaneously and instead should propagate as a pressure wave. A non-static situation is thus obtained. On assuming that : (i) the ether is conserved; (ii) the disturbed motion of the ether, i.e. its motion *with respect to the "macro-ether"* which defines the inertial frame, obeys Newton's second law; (iii) the disturbance in the fields ($p_{e1}$ for ether pressure and $\mathbf{v}_{e1}$ for ether velocity) is small, the latter being assessed with respect to the "sound" velocity

$$c_e(\rho_e) = \sqrt{\frac{dp_e}{d\rho_e}} \quad ; \quad (7)$$

and (iv) the undisturbed fields $p_{e0}$ and $\rho_{e0}$ of a local decomposition

$$p_e = p_{e0} + p_{e1} \ , \ \ p_{e1} \ll p_{e0} \ \text{ and } \ \partial p_{e0}/\partial t \ll \partial p_{e1}/\partial t \quad (7bis)$$

obey Eq. (6), it has been proved in [8a] that the general, non-static situation is governed by the equation :

$$\Delta p_e - \frac{1}{c_e^{\,2}} \frac{\partial^{\,2} p_e}{\partial t^2} = 4\pi \, G \, \rho \, \rho_e \quad . \quad (8)$$

Equation (8) clearly admits pressure waves, i.e. *gravitational waves* propagating at the velocity $c_e(\rho_e)$.

The "*macro-ether*" is a primary concept in the proposed theory. With respect to an arbitrary reference frame $R$, the velocity $\mathbf{v}_{e0}$ of the macro-ether is related to that of the microscopic ether (that which is thought to be concerned with particles and shorter-range fields) by an averaging over domains $\Omega$ such that the ether pressure is uniform in $\Omega$ (but varies over larger scales).

## 3. SPECIAL RELATIVITY AND ETHER THEORY

It is well-known that Lorentz and Poincaré, whose analyses were dependent on the assumption of the *rigid luminiferous ether*, derived some basic results (formulae) of special relativity (SR) before Einstein who's starting point was the *relativity principle*. Our point here is not to raise a discussion on the extent to which Lorentz and Poincaré did open SR : it is only to briefly recall why the whole of SR, in its usual formulation due to Einstein, can be consistently derived from the above ether assumption; this has been established by several physicists among which Builder [10] and Prokhovnik [11]. The crucial step is to recover the usual Lorentz transformation from the mere assumption (Fitzgerald and Lorentz) of an absolute contraction of all material objects when carried over from the rest frame $E$ of the ether to a frame $E_{\mathbf{u}}$ that moves with uniform and constant velocity $\mathbf{u}$ with respect to $E$. The contraction occurs on lines parallel to $\mathbf{u}$, but not on perpendicular lines. Its ratio $\beta(u)$ depends only on the modulus $u=|\mathbf{u}|$ (this results from the assumed isotropy of electromagnetic interaction in the frame $E$), *in a way that does not need to be assumed in advance* [8b]. Let $l$ be the length of a rod AB, when it was at rest in $E$, and $l'$ its length, evaluated from $E$, when this same rod is now at rest in $E_{\mathbf{u}}$ and makes, for $E$, the angle $\theta$ with $\mathbf{u}$. The assumed contraction implies that

$$l' = \frac{l\,\beta}{\sqrt{1 - (1-\beta^2)\,\sin^2\theta}} \quad . \qquad (9)$$

Consider AB (with a mirror at A perpendicular to AB and the like at B) as a "light clock", its time unit being the interval during which the light goes from A to B back and forth. To calculate the time unit $\Delta t_{\mathbf{u}}(\theta)$, *for $E$*, of this light clock as it is now in $E_{\mathbf{u}}$, we need only space and time measurements in $E$, so we can write if a light wave front going from A to B intersects the rod at M (and the point O being bound to $E$):

$$c_1 \frac{\mathbf{AB}}{AB} \equiv \frac{\mathrm{d}\,\mathbf{AM}}{\mathrm{d}t} = \frac{\mathrm{d}\,\mathbf{OM}}{\mathrm{d}t} - \frac{\mathrm{d}\,\mathbf{OA}}{\mathrm{d}t} = \frac{\mathrm{d}\,\mathbf{OM}}{\mathrm{d}t} - \mathbf{u} \quad , \quad \left|\frac{\mathrm{d}\,\mathbf{OM}}{\mathrm{d}t}\right| = c \quad , \quad (10)$$

whence $c_1 = \sqrt{c^2 - u^2\sin^2\theta}$ - $u\cos\theta$ and similarly $c_2 = \sqrt{c^2 - u^2\sin^2\theta}$ + $u\cos\theta$ in the **BA** direction. Therefore,

$$(\Delta t)_{\mathbf{u}}(\theta) = l'\left(\frac{1}{c_1} + \frac{1}{c_2}\right) = \frac{2\,l\,\beta}{c\left(1 - u^2/c^2\right)}\,\frac{\sqrt{1 - u^2\sin^2\theta/c^2}}{\sqrt{1 - (1-\beta^2)\sin^2\theta}} \quad . \qquad (11)$$

The "negative" Michelson-Morley experiment means exactly that this period does not depend on angle θ and thus is a constant in the moving frame $E_{\mathbf{u}}$. From (11), we see that this is true if and only if one has

$$\beta(u) = \sqrt{1 - \mathbf{u}^2 / c^2} \quad . \quad (12)$$

*The negative Michelson experiment means that light clocks are correct clocks, and this implies the Lorentz contraction*. Moreover, *the time unit of $E_{\mathbf{u}}$ is "dilated"* : $(\Delta t)_{\mathbf{u}} = \Delta t/\beta$. Prokhovnik shows precisely that these results, together with the standard synchronization of clocks, imply the kinematics of SR in Einstein's form, i.e. imply first the constancy of c (as this is measured with clocks and rods of any frame in uniform motion $E_{\mathbf{u}}$) and then the usual Lorentz transformation [11]. The difference with the usual version is in the perfectly clear interpretation of the space contraction and time dilation as absolute effects. The relativity of simultaneity is interpreted as an artefact due to the necessity of synchronizing clocks with light signals and to the anisotropic "true" propagation of light, Eq. (10), in a frame that moves with respect to the propagating medium $E$. The true simultaneity is defined with the time of $E$, and could take an operational significance if the velocity with respect to $E$ could be known.

Then, the dynamics of SR follow essentially from its kinematics and from the requirement to save Newton's second law in $E$ as well as at least a weak form of the principle of inertia (WPI) - namely that the momentum conservation for an isolated system of mass points pass from $E$ to $E_{\mathbf{u}}$ [8b,11]. One realizes first that this can be true only if the inertial mass depends on the relative velocity, m=m(v). On assuming that the m(v) dependence is the same in $E$ and in $E_{\mathbf{u}}$ , the WPI leads to the expression of m(v) and the kinetic energy T in SR :

$$m(v) = m(v=0)/\beta(v) \;,\quad T = (m(v)-m(0)) \, c^2 \quad . \quad (13)$$

Einstein's form of Newton's second law is thus obtained in $E$ :

$$\mathbf{F} = \frac{d\mathbf{P}}{dt} \quad,\quad \mathbf{P} = m(v) \, \mathbf{v} \quad,\quad (14)$$

and the requirement that it is Lorentz-covariant, i.e. holds true when passing from $E$ to $E_{\mathbf{u}}$ , determines the transformation of $\mathbf{F}$. The momentum conservation can be postulated also in the case of non-conserved rest masses (i.e. creation/annihilation processes), and this extension also passes from $E$ to $E_{\mathbf{u}}$ if and only if the total mass is conserved :

$$\sum_{i=1}^{N} m_i(v_i) = \sum_{j=1}^{N'} m'_j (v'_j). \quad (15)$$

Together with Eq. $(13)_2$, this gives the classical argument of SR for attributing to any particle the "rest energy" $m(0)c^2$ and identifying mass and energy, up to the factor $c^2$.

## 4. EQUIVALENCE PRINCIPLE AND ETHER THEORY

As shown by Fock [2] and Synge [16], Einstein's equivalence principle (EP) between effects of a gravitational field and inertial effects has to be used with some care; in particular, it is only valid in the infinitesimal, i.e. for uniform fields. Yet in a relativistic theory, the unifor-mity must be in space-time (i.e. the infinitesimal domain to consider is in space-time), thus the local equivalence should already apply with a uniform and *non-accelerated* motion. Hence a such motion should have "inertial" or pseudo-gravitational effects. This is readily verified, for in GR gravity manifests itself by effects on clocks and rods, and effects of this kind are indeed predicted by SR for non-accelerated motion; and again because the equivalence is for uniform fields, the metric effects of non-accelerated motion should be the essential ones when looking for the metric effects of gravity. This is apparent in Einstein's rotating disc. In this example, the metric effects are indeed assumed to depend on the local velocity, $v = r\omega$ ($\omega$ is the constant spin rate of the disc with respect to the inertial reference frame). Moreover, the pseudo-gravitational (acceleration) field has intensity $\mathbf{a} = r\omega^2 \, \mathbf{e}_r = \mathrm{grad}$ $(v^2/2)$; since the radial distance r is not fixed, the metric effects are bound in no way to the local intensity of the field, but instead to its potential (the same conclusion is drawn if one considers an accelerated translation). Thus, an analysis based on EP indicates unambiguously that the effects of gravitation on the space-time metric : (i) are bound to the gravity potential rather than to the field intensity, and (ii) should be deduced from the effects of non-accelerated motion. In the orthodox relativistic view, the second conclusion is difficult to accept, because no inertial frame is privileged there and the metric effects of non-accelerated motion are hence a kind of "parallax in space-time" rather than absolute ones, whereas the presence of a gravitational field is surely something "absolute", that cannot be removed by changing the reference frame - except, and for one part only, in the infinitesimal. The EP together with Einstein's lift and disc are often used in GR to introduce Einstein's idea that gravitation *is* (nothing else than) a curved space-time metric with (1,3) signature, the geodesic lines of which are the trajectories of free test particles. However, Einstein's field equations cannot be obtained directly as a modification of Newton's accounting for EP, and this might be related to the above-mentioned difficulty with the absolute effects of uniform motion.

In a theory based on ether, there is no logical difficulty in assigning absolute effects to uniform motion with respect to the ether. Furthermore, there is in the proposed ETG a basis for the EP when this is seen as a relationship between the metric effects of motion (depending only on the velocity, i.e. on the locally tangent uniform motion) and those of gravitation. Gravitation is for us a variation in the ether density $\rho_e$ , and a variation in the apparent ether density indeed occurs in uniform motion, due to the Lorentz contraction : for an observer having a constant velocity $\mathbf{u}$ with respect to $E$, a given volume $dV^0$ of ether has a greater volume $dV = dV^0/\beta(u)$, because his measuring rod is contracted in the ratio $\beta$ in the direction $\mathbf{u}$. The "mass" or rather the amount of ether is unchanged, for the mass increase with velocity concerns only material particles : the "mass" of the macro-ether is not a mass in the mechanical sense [8a]. Thus the apparent ether density is lowered, $\rho_{e\mathbf{u}} = \rho_e.\beta(u)$. This way of reasoning assumes that the moving observer can use the "true" simultaneity (of the frame $E$), which is true if he knows his velocity $\mathbf{u}$; with the simultaneity defined from clock synchronization in the frame $E_{\mathbf{u}}$, the apparent ether density would be $\rho'_{e\mathbf{u}} = \rho_e/\beta(u)$ since $dV'=dV^0.\beta(u)$ as deduced from standard Lorentz transformation. In any case, the metric effects of uniform motion are given by the ratio of an apparent ether density to that of the reference ether density, $\beta(u)=\rho_{e\mathbf{u}}/\rho_e = \rho_e/\rho'_{e\mathbf{u}} <1$. It is therefore natural to postulate that :

(A) *In a gravitation field, material objects are contracted, only in the direction of the field* $\mathbf{g}$=-grad $p_e/\rho_e$ , *in the ratio*

$$\beta_g = \rho_e / \rho_e^\infty < 1 , \qquad\qquad (16)$$

*where $\rho_e^\infty$ is the ether density at a point where no gravity is present, and the clock periods are dilated in the same ratio*.

This statement is made for objects and clocks bound with $E$; if this is not so, one has to combine the metric effects due to motion and gravitation. Due to the space-contraction of measuring rods in the direction $\mathbf{g}$, the physical space metric $\mathbf{g}$ in the frame $E$ becomes a Riemannian one. The contraction occurs with respect to an abstract Euclidean metric, which we assume that the macro-ether may be equipped with, as a 3-D manifold M : we assume that M is diffeomorphic to $\mathbf{R}^3$ ; let us precise that M is the same manifold at any time, since the body "macro-ether" (Sect. 2) is followed from its own rest frame. However, an infinity of different flat metrics may be defined on a such manifold [1]. It is thus more exact to formulate the above assumption in stating : *that space metric $\mathbf{g}^0$ (in the frame $E$) which is (uniquely) deduced from the physical one $\mathbf{g}$ by the dilation of length standards in the ratio $\rho_e^\infty/\rho_e$ >1 in the direction $\mathbf{g}$ , is a flat metric* [2]. Therefore, the value $\rho_e^\infty$ does not need to be reached : it is the unique value for which the assumption holds. There are three further arguments for assumption (A) :

(i) according to the ETG, the gravity acceleration $\mathbf{g}$ derives from the potential :

$$U = -c_e^2 \, Log(\rho_e/\rho_e^\infty), \qquad\qquad (17a)$$

---

[1] If any diffeomorphism of M onto $\mathbf{R}^3$, i.e. any everywhere regular coordinate system $(x^i)$, is given, one simply sets $g^0_{ij}(\mathbf{x})=\delta_{ij}$ for all $\mathbf{x}$; hence, even what is defined as "straight lines" is not fixed, and instead may be deformed very generally from one choice to another.

[2] However, there is still the possibility that $\mathbf{g}$ and $\mathbf{g}^0$ are related together by assumption (A) up to a time-dependent factor $\alpha(t)$ only : this will be important in cosmological problems, since it allows a global expansion of the ether relative to the physical space metric. Note also that $\mathbf{R}^3$ might be replaced by the sphere $\mathbf{S}^3$ (resp. the Lobatchevsky space $\mathbf{L}^3$) and hence M would be a space of constant positive (resp. negative) curvature equipped with its natural Riemannian metric $\mathbf{g}^0$ .

with = $c_e(\rho_e)$ (this is deduced from Eq. (5) after linearization of the $p_e$ - $\rho_e$ relationship around $\rho_e^{\infty}$ and holds thus either for a weak field or if the $p_e$ - $\rho_e$ relationship is linear); for a weak field:

$$U \approx c_e^2 \, (\rho_e^{\infty} - \rho_e)/ \, \rho_e^{\infty}, \qquad 1 - U/c_e^2 \approx \rho_e / \, \rho_e^{\infty} \qquad \qquad (17b)$$

Furthermore, in the theory of the fluid constitutive ether, the material particles are local flows in ether (vortex tori, or complexes of such flows) and therefore their speed is limited by the local "sound" velocity $c_e$ [13]. Since SR gives the other limit c, *one must have* $c_e \equiv c$, which implies that *the $p_e$-$\rho_e$ relationship is in fact linear*, $p_e = c^2 \, \rho_e$ . Now it is well-known that the EP implies that in a weak field any clock is slowed down in the ratio $\beta = 1 - U/c^2$ (this follows immediately from Einstein's analysis of the disc or lift, cf. supra), hence the EP together with Eq. (17) give Eq. (16) for weak fields.

(ii) It is easy to check that if one states assumption (A) without giving the ratio $\beta_g$ by Eq. (16), this latter (possibly combined with the $\alpha(t)$ factor, see above) is equivalent to say that, as evaluated with respect to the Euclidean metric $\mathbf{g}^0$, the ether density is uniform (in space). Since this metric is by hypothese bound to the ether, i.e. makes the ether a rigid body, it is only natural that it renders the ether density uniform and thus the unknown ratio $\beta_g$ must be given by Eq. (16). This result also confirms, if there could be any doubt, that in the modification of the ETG accounting for SR and EP, the equations will have to be understood as relative to the physical metric $\mathbf{g}$ instead of the abstract metric $\mathbf{g}^0$ , for the ETG assumes an ether compressibility $K\neq0$ (and now $K\equiv1/c^2$).

(iii) Let us investigate whether the ETG leaves the possibility of a Newton law in the frame *E*, in the presence of gravitation and Riemannian space-time (as already seen, the space and time standards must be the local, physical ones). The time-derivative of a vector (the momentum) may be consistently defined as a vector in the case of the Riemannian metric $\mathbf{g}$ on the manifold M, if $\mathbf{g}$ is time-independent [8b]. If $\mathbf{P}$ (e.g. the momentum of a mass point) is a vector attached to a moving point whose position $\mathbf{x}$ in M depends on the arbitrary time-coordinate t, one defines then its derivative (relative to *this* coordinate t) by

$$P^i_{\ ;\, t} = \frac{d \, P^i}{dt} + \Gamma^i_{j \, k} \, P^j \, v^k \quad , \qquad (18)$$

with $v^k = dx^k/dt$ and the $\Gamma$'s being the Christoffel symbols associated with the metric $\mathbf{g}=(g_{ij})$ in the considered coordinate system $(x^k)$; this defines a true vector (the components (18) are contravariant with respect to any transformation of *space* coordinates). Thus an exact Newton law can be defined if $\rho_e$ and hence $\mathbf{g}$ do not depend on time. The expression of the momentum is known from SR, $\mathbf{P}=m(u)\mathbf{u}$, the velocity $\mathbf{u}$ and its modulus u being defined with clocks and rods of the momentarily coincident point of M : $u^k = dx^k/dt_\mathbf{x} = v^k\, dt/dt_\mathbf{x}$ , $t_\mathbf{x}$ being the local time defined by assumption (A) : if $t = t_{\mathbf{x}_0}$ is the proper time of a fixed point $\mathbf{x}_0 \in$ M, one gets in general from Eq. (16) :

$$dt/dt_\mathbf{x} = dt_{\mathbf{x}_0}/dt_\mathbf{x} = \rho_e(\mathbf{x}_0,t)/\rho_e(\mathbf{x},t) = \rho_{e0}/\rho_e \ , \quad (19)$$

though in the present case $\rho_e$ does not depend on t. The gravitation force involves the gravity acceleration $\mathbf{g}$ (Eq. (5) where the gradient is defined with $\mathbf{g}$) and the passive gravitational mass $m^p$. According to the EP, in its weak form already true in Newtonian theory, $m^p$ equals the inertial mass, here m(u). Thus our modified Newton law writes :

$$\mathbf{F} \equiv \mathbf{F}_0 + m(u)\, \mathbf{g} = \frac{d}{dt_\mathbf{x}} \left( m(u) \frac{d\mathbf{x}}{dt_\mathbf{x}} \right) , \quad \frac{d}{dt_\mathbf{x}} \equiv \frac{\rho_{e0}}{\rho_e} \frac{d}{dt_{\mathbf{x}_0}} \quad , \quad (20)$$

where $\mathbf{F}_0$ is the non-gravitational force and the time derivative of the momentum is calculated with Eq. (18); equation (20) is invariant under the change of the reference point bound with ether, $\mathbf{x}_0$. Now we have the following result [8b] :

(G) *if the gravitation field is constant in the frame* $E$*, then the solution trajectories of the modified Newton law with purely gravitational force, Eq. (20) with* $\mathbf{F}_0 = 0$*, are geodesic lines of the space-time metric deduced by assumption* (A); the line element of this metric $\gamma$ writes in the frame $E$ :

$$ds^2 = c^2\, d\tau^2 = (\rho_e/\rho_{e0})^2\, (dx^0)^2 - dl^2 \ , \qquad (21)$$

where $\tau$ is the proper time of the test particle, $x^0 = ct$ ($t = t_{\mathbf{x}_0}$) and $dl^2$ is the space line element calculated with the space metric $\mathbf{g}$ in the frame $E$ :

$$dl^2 = u^2\, dt_\mathbf{x}^2 \ , \quad u^2 = \mathbf{g}(\mathbf{u},\mathbf{u}) = g_{ij}\, u^i\, u^j \ . \qquad (22)$$

Thus, the modified Newton law (20), which is not even Lorentz-covariant, implies the geodesic characterization of motion which is generally covariant. We assume here, by induction from the result (G), that free test particles also follow the geodesic lines of $\gamma$ if $\rho_e$ and $\mathbf{g}$ are time-dependent, in which case we have not defined a Newton law. However, if the time variation of $\rho_e$ is small in such a way that the corresponding terms are negligible in the equation of space-time geodesics (or more generally in the expression of the four-acceleration of any test particle, "free or not"), then an approximate Newton law (20) is obtained by still using the definition (18) for the derivative of the momentum. Note that, due to Eq. (21), the components $\gamma_{0i}$ (i=1,3) of $\gamma$ are always nil in the frame $E$. In particular, a *constant* (time-independent) $\rho_e$ corresponds to what is called a *static* field in GR.

5. FIELD EQUATION, NEWTONIAN LIMIT AND INERTIAL FRAMES

The result (G) does not depend on the field equation. To adapt this latter to the situation where the metric modifications due to motion and gravitation are accounted, we first note that, in the static case, the whole of the arguments in Sect. 2 may be repeated to obtain Eqs. (5) and (6), but this time with the Riemannian space metric **g** (the requirement that Newtonian theory must be recovered for an incompressible ether remains as it stands since, from Eq. (16), the Euclidean metric is regained simultaneously with Poisson's equation as the compressibility tends towards zero). In the non-static case, the reasonment, detailed in [8a] and based on the assumptions (i) to (iv) in Sect. 2 here, is not so easy to modify in a clear-cut way, in so far as we have no Newton law any more. However, the same equation (8) should be obtained in terms of the physical space and time metrics in the frame $E$ :

$$\Delta_{\mathbf{g}} \, p_e - \frac{1}{c^2} \frac{\partial^2 p_e}{\partial t_{\mathbf{x}}^2} = 4\pi \, G \, \rho \, \rho_e \quad . \ (23)$$

Indeed, the proposed approach leads unambiguously to this equation in the static case, and Eq.(23) is merely the natural expression of Eq. (8) once it has been recognized that the non-uniform *physical* space and time metrics are relevant. Thus one is strongly inclined to assume Eq. (23) - and so we do. We emphasize, however, that in (23) the Laplace operator $\Delta$ is defined with the physical space metric **g** which is curved and whose expression in terms of the Euclidean metric $\mathbf{g}^0$ bound to M, *depends on the unknown* $p_e$ through assumption (A); in the same way, the local time $t_{\mathbf{x}}$ depends on $p_e$ or $\rho_e$ (Eq. (19)). Hence the left-hand side of Eq.(23) is not that of d'Alembert's classical wave equation and instead depends non-linearly on $p_e$ . It is also worth to note that the operator on the left of Eq. (23), which is linear if **g** or $p_e$ is considered given, but which is here precisely applied to $p_e$ , is *not* even the modified d'Alembert or    operator of GR, which is defined from the Riemannian space-time metric **γ** by

$\phi = \text{div}_{\boldsymbol{\gamma}} \, (\text{grad}_{\boldsymbol{\gamma}} \, \phi) = (1/\sqrt{|\gamma|}) \, (\partial/\partial x^{\alpha})(\sqrt{|\gamma|} \, \gamma^{\alpha\beta} \, \partial\phi/\partial x^{\beta}) \ , \ (24)$

with $\gamma = \det (\gamma_{\alpha\beta})$. In fact, the Laplace operator in (23) is defined from the same formula by substituting the space metric **g** for the space-time metric **γ** , and thus with summation on latine indices i and j (from 1 to 3) instead of that one on greek indices $\alpha$ and $\beta$ (from 0 to 3) in Eq. (24) :
$\Delta_{\mathbf{g}}\phi = \text{div}_{\mathbf{g}} \, (\text{grad}_{\mathbf{g}} \, \phi) = (1/|g|) \, (\partial/\partial x^i)( \, \sqrt{|g|} \, g^{ij} \, \partial\phi/\partial x^j) \ . \ (25)$

Due to this non-linearity, Eq. (23) contains features of classical equations for intense wave propagation (e.g. those of gas dynamics), in particular it contains the possibility of shock waves (Sect. 8). There is another non-linearity in Eq. (23), since its right-hand side already contains a non-linear dependence of $p_e$ on $\rho$; in a sense, this right-hand side implies that the gravitation field, characterized by $p_e$ or $\rho_e$, is a source to itself. Moreover, the mass-energy equivalence of SR has to be accounted for in the definition of the mass distribution $\rho$. Thus $\rho$ *in Eq. (23) is defined to be the density of mass-energy in the frame E*. As imposed by the inclusion in $\rho$ of the energy of light-like particles (divided by $c^2$), the mass of ordinary particles involved in $\rho$ is not the rest-mass but the velocity-dependent one, which is to say that the Newtonian equality of active and passive gravitational mass is saved. It implies that the mass equivalent of the ordinary, kinetic pressure p in macroscopic bodies is included in $\rho$, in other words pressure does contribute to gravity in the ETG, as in GR.

It appears that *Eq. (23) has no covariance property*, but this is a priori tenable since all quantities are understood in the frame *E*, which according to the theory is strongly privileged. The next thing to do is thus to show that the Newtonian approximation, which for most of the reliable observations is very accurate and which implies the Galilean invariance, is recovered for weak and slowly varying fields in this modified ETG. In GR, such fields are characterized by the property that one may choose the coordinate system $(x^\alpha)$ so that the space-time metric **γ** differs from the Galilean one **η** (with $\eta_{\alpha\beta} = \delta_{\alpha\beta}$ if $\alpha=0$ and $\eta_{\alpha\beta} = -\delta_{\alpha\beta}$ if $\alpha=1,2,3$) only by evanescent terms : $\gamma_{\alpha\beta} = \eta_{\alpha\beta} + h_{\alpha\beta}$ and moreover the derivatives $\partial h/\partial x^0$ , with $x^0=ct$, are negligible as compared with the space derivatives $\partial h/\partial x^i$. Similarly, we admit that the field $p_e$ has the form

$$p_e = p_e^\infty(1-U/c^2), \ \ U/c^2 \ll 1 , \quad (26)$$

and we will neglect the derivatives of the kind $\partial p_e/\partial x^0$ with respect to the corresponding space derivatives. Eq. (26) obviously means that the gravitation field in the sense of the ETG (**g** in Eq. (5)) is weak; moreover, the deviations from the Euclidean space metric $\mathbf{g}^0$ and the uniform time t are directly given by the ratio $p_e/p_e^\infty$ (Eq. (16)). If we select a Cartesian coordinate system $(x^i)$ on M, i.e. $g^0_{ij}=\delta_{ij}$ , it follows thus from Eq. (26) that the Laplace operator $\Delta$ according to **g** (Eq. (25)) will differ negligibly from that one according to $\mathbf{g}^0$, $\Delta_0 \phi=\partial^2\phi/\partial x^i\partial x^i$, and that the derivatives with respect to the local time $t_\mathbf{x}$ may be replaced by ones with respect to the time t. If we account further for the assumption of a slowly varying field, we rewrite Eq. (23) in the form

$$\Delta_0 \, p_e = 4\pi G \, \rho \, \rho_e \approx 4\pi G \, \rho \, \rho_e^\infty. \quad (27)$$

Now with $p_e = c^2 \rho_e$, the field $\mathbf{g} = -\,(\text{grad } p_e)/\rho_e$ is, by Eq. (26), equivalent to $-(\text{grad } p_e)/\rho_e^\infty$ , thus we recover the incompressible case with Euclidean metric, i.e. the Newtonian equation for the gravitation field :

$$\Delta_0 \, U = -\, 4\pi G \, \rho \ . \quad (28)$$

In other words U in Eq. (26) is equivalent to the Newtonian potential. It may be worth to give a simple reason why the spatial derivatives should dominate the $x^0$-derivatives in the Newtonian limit. The Newtonian potential is the convolution product : $U=G\rho*(1/r)$ with r the spatial distance, so that:

grad $U = G \rho * \text{grad}(1/r)$ ,

$\partial U/\partial x^0 = (1/c) \, \partial U/\partial t = (1/c)G\rho * \partial (1/r)/ \partial t = (1/c)G\rho * [(\partial \mathbf{r}/\partial t).\text{grad}(1/r)]$ , (29)

and $-\partial \mathbf{r}/\partial t$ in (29) is the velocity of the current point $\mathbf{y}$ in the set of massive bodies, with respect to the point $\mathbf{x}$ where the field is computed. Thus for low velocities one has $|\partial \mathbf{r}/\partial t| << c$ for any $\mathbf{y}$ and $\partial U/\partial x^0$ is indeed negligible with respect to grad U.

The Newtonian field equation (28) may be written in any reference frame (as far as the space metric is considered Euclidean) and thus does not define the inertial frames. These are obtained from the Galilean invariance of the non-relativistic Newton law; in the ETG, the latter is derived in exactly the same way as in GR, from Einstein's geodesic assumption for weak and slowly varying fields and small velocities. In this derivation, the $x^0$-derivatives of the metric are again neglected. Thus the approximate inertial frames that appear for weak and slowly varying fields are such that the corresponding expression of the metric is close to be static, as is confirmed by the result (G). The basic assumption (A) (the gravitational space contraction in the ratio $\beta$ of Eq. (16)) relates the physical space metric in the frame $E$ to the abstract Euclidean metric which makes the macro-ether a rigid body. Thus, in stating assumption (A) we force the ether to remain the absolute space also in the modified version of the proposed theory; moreover, the iterative approximation of the proposed ETG confirms that the inertial frames of the first (Newtonian) approximation are in uniform motion with respect to the macro-ether (this iterative scheme is not presented in this paper).

Note that in the static case ($\rho_e$ time-independent), Eq. (23) *implies* that $\rho$ also does not depend on t in the frame $E$; since in that case the modified Newton law holds, there can be only one body (any other body would fall, in contradiction with the constancy of $\rho$) and this body must be in equilibrium and at rest in the frame $E$. Thus in the static case there is no relative motion of ether and matter at the macro-scale. There is another case where the proposed ETG leads to this "total dragging" situation. We have postulated a *constitutive* ether: according to this, material (elementary) particles would be local flows in ether [13]. If one accepts this, then the macroscopic motion of the ether (defining the frame $E$) must include the motion of matter, since the former motion is a weighting of the motions of the ether "constituting the particles" and of the ether "outside the particles". As realized above, the macro-ether remains the absolute space, hence the particles cannot drag all the surrounding fluid. However, the mean motion of ether is directly influenced by the motion of particles. Since the ether compressibility is assumed to be extremely low ($K=1/c^2$), the "amount" of ether inside the particles contained in a volume $\Omega$, $m_e^{\text{part}} = \int_{\text{particles in } \Omega} \rho_e \, dV$, is nearly proportional to the volume they occupy, $V^{\text{part}} = \int_{\text{particles in } \Omega} \, dV$. Hence, at ordinary densities where the particles occupy only a very small volume fraction, the direct (kinematical) influence of the motion of matter on the macroscopic motion of ether is likely to be very small. But consider the case of a body in advanced *gravitational collapse*, where the material density would be higher than the nuclear density; in that case, it is natural to admit that the particles are "contiguous to each other", i.e. that all ether is involved in material particles : the macro-ether is automatically comoving with matter (see Sect. 8 for an application).

## 6. THE MODIFIED NEWTON LAW IMPLIES GEODESIC MOTION OF LIGHT-LIKE PARTICLES

A basic prediction of GR, also found with the ETG (Eq. (19)), is that clocks are slowed down in a gravitation field. One empirical confirmation is the observed gravitational red shift of electromagnetic spectra. To deduce the red shift from the slowing down of clocks, the frequency $\nu_0$ of any monochromatic wave is considered constant throughout its light path, when expressed in terms of the time coordinate t - in the case of a stationary field where the space-time metric ($\gamma_{\alpha\beta}$) does not depend on t. This constancy of $\nu_0$ in turn is derived from the wave equation $\quad \phi=0$ with $\phi=\exp(i\psi)$, $\psi=\omega(x^1,x^2,x^3)t + f(x^1,x^2,x^3)$ and $\quad \phi$ from Eq. (24) : Laue proved this for a static field [17] but his arguments extend to a stationary one, in the way outlined in the treatise [18]. Laue [17] also deduced from the same way of reasoning that light rays follow the null geodesic lines of $\boldsymbol{\gamma}$ . These arguments also apply in the ETG, since they depend only on the equation $\quad \phi=0$, in the limit of geometrical optics. It is, however, interesting to show that in the ETG the constancy of $\nu_0$ as well as the geodesic propagation can be deduced also from the modified Newton law (20) with the mass of the energy E in the place of the inertial mass m(u) :

$$ E\, \mathbf{g} = \frac{d}{dt_{\mathbf{x}}} \left( E\, \frac{d\mathbf{x}}{dt_{\mathbf{x}}} \right) , \quad E = h\nu \quad (30) $$

in which $\nu$ is the frequency as measured with the local (momentarily coinciding) clock, $\nu=dn/dt_{\mathbf{x}}$, as imposed by the local nature of quantum phenomena (h is Planck's constant and n is the number of oscillations). The constancy of $\nu_0 = dn/dt = (p_e/p_{e0})\nu$ means that the energy of the photon decreases if it moves off from the attracting body, in such a way that :

$$ E\, p_e = E_0\, p_{e0} = Const. \quad \text{along the light path.} \quad (31) $$

We first prove that for a static field ($p_e$ time-independent), (31) follows from (30) and the fact that, as measured with physical clocks and rods (bound to *E*), the modulus of the velocity of the photon is always equal to c :

$$ \left( \frac{d\mathbf{x}}{dt_{\mathbf{x}}} \right)^2 = c^2 , \quad \frac{d}{dt_{\mathbf{x}}}\left[ \left( \frac{d\mathbf{x}}{dt_{\mathbf{x}}} \right)^2 \right] = 2 \left( \frac{d\mathbf{x}}{dt_{\mathbf{x}}} \right) \cdot \frac{d}{dt_{\mathbf{x}}} \left( \frac{d\mathbf{x}}{dt_{\mathbf{x}}} \right) = 0 \quad (32) $$

(the usual notations for scalar products and squares, $\mathbf{u.v}$ and $\mathbf{u}^2$, refer to the curved space metric $\mathbf{g}$). Combining $(30)_1$ and $(32)_2$ we get :

$$E\, \mathbf{g}.\frac{d\mathbf{x}}{dt_\mathbf{x}} = \frac{dE}{dt_\mathbf{x}}\left(\frac{d\mathbf{x}}{dt_\mathbf{x}}\right)^2 = c^2\, \frac{dE}{dt_\mathbf{x}}\ , \qquad (33)$$

but according to Eq. (5) where the grad operator refers to the metric $\mathbf{g}$ and with $p_e = \rho_e c^2$, we have

$\mathbf{g.dx} = -\,(\text{grad } p_e . d\mathbf{x})/\,\rho_e = -\,c^2\, dp_e/p_e$ .

Therefore, (33) may be rewritten as

$-\,E\ c^2\, dp_e/p_e = c^2\, dE\ ,\quad (35)$

which means that $d(Ep_e) = 0$ along the path, i.e. Eq. (31) is obtained.

Now we prove that Eqs. (30) and (31) imply that the photon moves along a null geodesic line of $\boldsymbol{\gamma}$ (*during this proof, we omit the index* e *in* p *and* $\rho$ *for simplicity*). Let $\mathbf{x}_0$ be any fixed point in M (i.e. $\mathbf{x}_0$ is at rest in the ether), outside the gravitation field (i.e. far enough). We introduce the universal time coordinate $x^0 = ct$ with $t = t_{\mathbf{x}_0}$. Whether the field p depends on t or not, we may adopt, at any time t, a space coordinate system $(x^i)$ on M such that p=Const. is equivalent to $x^1$=Const., and such that the Euclidean metric $\mathbf{g}^0$ is diagonal in the natural basis $(\mathbf{e}_i)$ :

$(g^0_{ij}) = \text{diag}\ (a_1^{\ 0},\, a_2^{\ 0},\, a_3^{\ 0})$.

Then, assumption (A) and Eq. (16) with $p = \rho\, c^2$ mean that $\mathbf{g}$ also is diagonal, and that:

$$(g_{ij}) = \text{diag}\left(\left(\frac{p^\infty}{p}\right)^2 a_1^{\ 0}, a_2^{\ 0}, a_3^{\ 0}\right) \equiv \text{diag}\left(a_i\right)_{1 \le i \le 3}\ , \qquad (36)$$

with $p^\infty = p(\mathbf{x}_0)$. In the case where such an "isopotential" coordinate system is bound to the frame $E$, the space-time metric (21) writes in the local coordinates $(x^\alpha)$ $(\alpha=0,3)$ of the space-time $\mathbf{R}\times M$:

$$(\gamma_{\alpha\beta}) = \text{diag}\left(\left(\frac{p}{p^\infty}\right)^2, -\left(\frac{p}{p^\infty}\right)^2 a_1^{\ 0}, -a_2^{\ 0}, -a_3^{\ 0}\right) \equiv \text{diag}\left(b_\alpha\right)_{0 \le \alpha \le 3}\ , \qquad (37)$$

with $b_i = -a_i$ for $1 \le i \le 3$. This case includes the static case ($\partial p/\partial t = 0$) and that of spherical symmetry around a fixed point in ether. The Christoffel symbols $\Gamma$ and $\Gamma'$ of the metrics $\mathbf{g}$ and $\boldsymbol{\gamma}$, respectively, have been calculated in the previous work [8b]. The first-kind symbols verify :

$\Gamma'_{ijk} = - \Gamma_{ijk}$  (i,j,k = 1,2,3); (38)

$\Gamma'_{\alpha\alpha\alpha} = b_{\alpha,\alpha}/2,$    $\Gamma'_{\alpha\beta\gamma} = 0$  if $\alpha \neq \beta \neq \gamma \neq \alpha,$ (39)

$\qquad \Gamma'_{\alpha\alpha\beta} = - \Gamma'_{\beta\alpha\alpha} = \gamma_{\alpha\alpha,\beta}/2 = b_{\alpha,\beta}/2$  if $\alpha \neq \beta;$

$- \Gamma'_{0\alpha\alpha} = \Gamma'_{\alpha\alpha 0} = \Gamma'_{\alpha 0\alpha} = b_{\alpha,0}/2, \Gamma'_{\alpha 0\beta} = \Gamma'_{\alpha\beta 0} = \Gamma'_{0\alpha\beta} = 0$ if ($\alpha \neq 0$ and $\beta \neq 0$ and $\alpha \neq \beta$), (40)

$\Gamma'_{00\alpha} = \Gamma'_{0\alpha 0} = - \Gamma'_{\alpha 00} = b_{0,\alpha}/2 = p\, p_{,\alpha}/(p^{\infty})^2$   . (41)

The second-kind Christoffel's are defined by

$\quad \Gamma^i_{jk} = g^{il}\, \Gamma_{ljk} = a_i^{-1}\, \Gamma_{ijk}$  ,  $\Gamma'^{\alpha}_{\beta\gamma} = b_{\alpha}^{-1}\, \Gamma'_{\alpha\beta\gamma}$  . (42)

Since $b_i = - a_i$ for $1 \leq i \leq 3$, we thus have from (38):

$\Gamma^i_{jk} = \Gamma'^i_{jk}$   (i,j,k = 1,3) (43)

The geodesic equation for $\alpha=0$, writes in view of (40) and (42) :

$\quad G^0 = \dfrac{d^2 x^0}{d\lambda^2} + \Gamma'^0_{\alpha\beta} \dfrac{dx^{\alpha}}{d\lambda}\, \dfrac{dx^{\beta}}{d\lambda} = \dfrac{d}{d\lambda}\left(\dfrac{dx^0}{d\lambda}\right) + 2\,\Gamma'^0_{i0}\, \dfrac{dx^i}{d\lambda}\, \dfrac{dx^0}{d\lambda} = 0$ , (44)

and since we expect a null geodesic, this equation serves to determine the time dependence of the parameter $\lambda$ : $y=dx^0/d\lambda$. It comes from (41) and (42) that :

$\Gamma'^0_{i0} = b_0^{-1}\, \Gamma'_{0\,i0} = p_{,i} / p$   (i=1,3) (45)

which is nil, unless if i=1. We get from (44) and (45) :

$G^0 = \dfrac{d}{d\lambda}\left(\dfrac{dx^0}{d\lambda}\right) + 2\,\dfrac{p_{,1}}{p}\dfrac{dx^1}{d\lambda}\,\dfrac{dx^0}{d\lambda} = \dfrac{dx^0}{d\lambda}\dfrac{d}{dx^0}\left(\dfrac{dx^0}{d\lambda}\right) + \dfrac{2}{p}\dfrac{dp}{dx^0}\left(\dfrac{dx^0}{d\lambda}\right)^2 = 0$  , (46)

in which $dp/dx^0$ is due only to the spatial variation of p along the trajectory. Eq. (46) gives

$dy/y + 2\, dp/p = 0$ ,  $y = dt/d\lambda = 1/p^2$  , (47)

in which $dt/d\lambda$ has been substituted for $dx^0/d\lambda$ , since $\lambda$ may be replaced by a proportional parameter. By (43), the left-hand side of the geodesic equation writes for i=1,3 :

$$G^i = \frac{d^2 x^i}{d\lambda^2} + \Gamma'^i_{\alpha\beta} \frac{dx^\alpha}{d\lambda} \frac{dx^\beta}{d\lambda} = \frac{d^2 x^i}{d\lambda^2} + \Gamma'^i_{jk} \frac{dx^j}{d\lambda} \frac{dx^k}{d\lambda} + \Gamma'^i_{00} \left(\frac{dx^0}{d\lambda}\right)^2 + 2 \Gamma'^i_{0k} \frac{dx^0}{d\lambda} \frac{dx^k}{d\lambda}$$

$$(48)$$

Now by (31) and (19), we write the neo-Newtonian law (30) as:

$$\frac{1}{p} \mathbf{g} = \frac{\left(p^\infty\right)^2}{p} \frac{d}{dt}\left(\frac{1}{p^2} \frac{d\mathbf{x}}{dt}\right), \quad (49)$$

whence by $(47)_2$ and (18) :

$$\mathbf{g} = \left(p^\infty\right)^2 \frac{d}{dt}\left(\frac{d\mathbf{x}}{d\lambda}\right) = \left(p^\infty\right)^2 p^2 \frac{d}{d\lambda}\left(\frac{d\mathbf{x}}{d\lambda}\right) = \left(p^\infty\right)^2 p^2 \left(\frac{d^2 x^i}{d\lambda^2} + \Gamma^i_{jk} \frac{dx^j}{d\lambda} \frac{dx^k}{d\lambda}\right)\mathbf{e}_i , \qquad (50)$$

thus the two first terms in the last member of (48) represent $g^i/[p^2 (p^\infty)^2]$. From (5) and (36) :

$$g^i = -\frac{c^2}{p} g^{ij} p_{,j} = -\frac{c^2}{p} a_i^{-1} p_{,i} = \begin{cases} 0 & \text{if } i \neq 1 \\ -c^2 p p_{,1} \big/ \left(\left(p^\infty\right)^2 a_1^0\right) & \text{if } i = 1 \end{cases} \quad (51)$$

Hence, by (40), the geodesic equation $G^i=0$ is also verified for i=2,3, and the last is

$$G^1 = -c^2 \frac{p_{,1}}{a_1^0 \, p\left(p^\infty\right)^4} + \Gamma'^1_{00}\left(\frac{dx^0}{d\lambda}\right)^2 . \qquad (52)$$

But from (37), (41), (42) and (47), we get:

$$\Gamma'^1_{00} = \frac{p^3 \, p_{,1}}{a_1^0 \, \left(p^\infty\right)^4} , \quad \left(\frac{dx^0}{d\lambda}\right)^2 = \frac{c^2}{p^4} . \qquad (53)$$

Hence the geodesic equation for photons (or other particles moving at the velocity of light) is proved independently of any field equation, from Newton's second law in the frame *E*. Like for ordinary particles, we assume here, by induction from this result obtained for a static field, that the geodesic motion holds true in the general case where a Newton law has not been defined.

## 7. SPHERICAL SYMMETRY : STATIC SOLUTION AND HYDROSTATIC EQUILIBRIUM

A spherical gravitation field has to be defined as the case where $p_e$ or $\rho_e$ depend only on the radial distance r (as measured with the Euclidean metric **g**$^0$ making the macro-ether a rigid body M) and the time t (as measured with one clock at a given point $\mathbf{x}_0 \in$ M, i.e. bound to *E*). Then one "isopotential" coordinate system, in which the metrics **g** and **γ** are given by (36) and (37), is the spherical coordinate system (r,θ,φ). Thus the field equation (23) writes, in view of (19) and (25) :

$$\frac{1}{r^2}\frac{p_e}{p_e{}^\infty}\frac{\partial}{\partial r}\left(r^2\frac{p_e}{p_e{}^\infty}\frac{\partial p_e}{\partial r}\right) - \frac{1}{c^2}\frac{p_e{}^\infty}{p_e}\frac{\partial}{\partial t}\left(\frac{p_e{}^\infty}{p_e}\frac{\partial p_e}{\partial t}\right) = A\,p_e\,\rho, \; A = \frac{4\pi G}{c^2} \;, \qquad (54)$$

in which one may moreover assume that $p_e{}^\infty$ does not depend on t, i.e. a static situation at infinity (or far enough from the massive body). Equation (54) implies that the density of mass-energy ρ also depends on t and r only. The *static* spherical solution ($p_e$ and *hence* ρ independent of t) follows easily from (54) and was given for a general distribution ρ(r) with finite integral M=M(r=∞)<∞ in the previous work [8b] (no confusion can occur between the mass M and the 3D-manifold M consisting of the macro-ether):

$$p_e{}^2 = -\frac{C}{2\pi r} - 2\frac{(p_e{}^\infty)^2}{c^2}U_N(r) + (p_e{}^\infty)^2 \;, \; U_N(r) = \int_r^\infty \frac{GM(u)}{u^2}du \;, \qquad (55)$$

where $M(r) = 4\pi \int_0^r u^2\,\rho(u)\,du$; the integration constant C is non-positive, and negative only in the unphysical case where an (1/r$^2$)-repulsion occurs for r→0 and at the same time the field **g** is equivalent to GM$^a$/r$^2$ with M$^a$<M at large r. In the physical case C=0, the space-time line element is thus, by Eq. (37) (with $a_1^0 = 1, a_2^0 = r^2, a_3^0 = r^2 \sin^2\theta$ ):

$$ds^2 = \left(1 - \frac{2U_N(r)}{c^2}\right)c^2 dt^2 - \frac{1}{1 - \frac{2U_N(r)}{c^2}}dr^2 - r^2 d\Omega^2 \;, d\Omega^2 = d\theta^2 + \sin^2\theta\,d\phi^2 \qquad (56)$$

and moreover the field **g** given by Eq. (5) (where the grad operator refers to the curved metric **g**), i.e. the gravity acceleration field in the sense of the modified Newton law (20) or (30), is the exact Newtonian field

**g** = - G M(r) **e**$_r$/r$^2$   (57)

(this is proved in [8b]). We emphasize that all of this is true for a general function ρ(r); in particular, if ρ(r)=0 for r>R, it is valid for r≤R as well as for r≥R. In that case, the metric (56) for r≥R is the exact Schwarzschild exterior metric ($U_N(r)$=GM/r). However, the metric for r<R is *not* that predicted by GR. E.g. in the case of a uniform density, ρ(r)≡ρ₀ for r<R, the term $U_N(r)$ in (56) is given, according to Eq. $(55)_2$, by

$$U_N(r) = \frac{GM}{R}\left(\frac{3}{2} - \frac{r^2}{2R^2}\right), \ 0 \le r \le R, \qquad (58)$$

whereas Schwarzschild's interior solution is :

$$ds^2 = \left(A - B\sqrt{1 - \frac{r^2}{R'^2}}\right)^2 c^2\, dt^2 - \frac{1}{1 - \frac{r^2}{R'^2}}\, dr^2 - r^2\, d\Omega^2, \qquad (59)$$

$$1/R'^2 = 8\,\pi\, G\, \rho_0\, /(3c^2)$$

with A>0 and B=1/2 (see e.g. [19]). The main difference between the interior solutions according to GR and the ETG is the following one : according to the ETG, the ratio

$$d l\, /\, d l^0 = \sqrt{g_{r\,r}} = \sqrt{(-\gamma_{r\,r})} > 1$$

between the radial distances as evaluated with the physical metric **g** and the Euclidean one **g**⁰ (which is equal to **g** at infinity) continues to increase as r goes from R to 0, whereas it decreases towards 1 for GR. *The result of the ETG is in agreement with what is found from the EP*, that the metric effects of gravity are given (for weak fields) by the Newtonian potential (Sect. 4), since $U_N(r)$ in Eqs. (55),(56) and (58) turns out to be nothing else than the Newtonian potential - however, no restriction to the field strength is imposed here. Another difference with GR is that, independently of the state equation, *the radius R of the body cannot be smaller than the Schwarzschild radius* $r_S = 2\ GM/c^2$, for otherwise Eq. (55) (with C=0, and $U_N(r)$=GM/r for r≥R) would lead to a negative square $p_e^2$. The number R is the radius of the body as evaluated with the Euclidean metric **g**⁰, hence the physical radius is greater. On the other hand, the mass-energy density ρ is evaluated with the physical metric **g** , hence M= ∫ρ√g⁰ dr dθ dφ = ∫ρ dV⁰ is smaller than the mass of the whole energy creating the field, M' = ∫ρ dV = ∫ρ√g dr dθ dφ , since

$$\sqrt{g^0}/\sqrt{g} = dV^0\ /dV = \sqrt{(1-2\ U_N(r)/c^2)} < 1.$$

The same occurs in GR, where M is interpreted as the sum of M' and the negative energy of the gravitational field [4]. A discussion of the gravitational energy in the present theory will be given elsewhere.

Since pressure does contribute to gravity in the ETG as in GR, one may expect that, as in GR, the static equilibrium is impossible if the integral M is too large : first, SR at the local scale and the definition of the energy-momentum tensor **T** hold true in the ETG; second, if one admits the geodesic formulation of motion in the general case (as is done in this paper), then the "conservation equation" (CE), $\text{div}_\gamma \mathbf{T} = 0$, can be used in the ETG (the geodesic motion of free particles is equivalent to the CE for **T** with "dust" matter; the essential difference in the ETG is that geodesic motion and the CE are not a consequence of the field equation for gravity). In that case, the static equilibrium writes exactly as in GR (see e.g. Weinberg [4]). For the case of spherical symmetry, it gives using the static solution (56)-(57) :

$$ -\frac{1}{(\rho + p/c^2)}\ \frac{dp}{dr}\ =\ \frac{G\ M(r)}{r^2\ (1 - 2\ U_N(r)/c^2)}\quad,\quad(60) $$

which is not identical to the Oppenheimer-Volkoff equation. However, it may be shown that it gives the same qualitative result, i.e. the existence of a superior limit for the mass M.

## 8. SPHERICAL SYMMETRY : GRAVITATIONAL COLLAPSE AND GRAVITATION WAVES

From above, we take it for granted that a very massive object will necessarily undergo a dynamic transformation, either an explosion or, in the case where the (ordinary) pressure p in the body is unable to balance the gravitation force, an *implosion*. Since pressure increases gravity also in the ETG, we expect that, as in GR, some qualitative features of the implosion are rather resistant to the change of the unknown state equation and therefore can be found from the analysis of the oversimplified situation of dust matter. Thus, the pressure p is *neglected*. The particles of the collapsing body are hence free and they have a geodesic motion. Furthermore, we assume here that the body, made of freely moving dust particles, is *comoving with the macro-ether* (see the end of Sect. 5 : the assumption of a constitutive ether leads naturally to this hypothese for very high densities such that elementary particles are more or less contiguous to each other). We thus write in any coordinates bound to the frame *E* :

$$ \frac{d^2\ x^\alpha}{ds^2}\ +\ \Gamma'^{\alpha}_{\beta\gamma}\frac{dx^\beta}{ds}\ \frac{dx^\gamma}{ds}\ =\ \frac{d^2\ x^\alpha}{ds^2}\ +\ \Gamma'^{\alpha}_{00}\left(\frac{dx^0}{ds}\right)^2\ =\ 0\ ,\quad(61) $$

which is equivalent to

$$ \frac{d^2\ x^0}{ds^2}\ +\ \Gamma'^{0}_{00}\left(\frac{dx^0}{ds}\right)^2\ =\ 0\quad\text{and}\quad\Gamma'^{i}_{00} = 0\ (i=1,3)\ .\quad(62) $$

For the case of spherical symmetry around the center O of the massive body which is assumed to be comoving with ether, the spherical coordinates $x^1 = r$, $x^2 = \theta$, $x^3 = \phi$ of the Euclidean space M (the body "macro-ether"), with origin O, are indeed bound to the frame *E* and moreover are such that $p_e$ does not depend on $x^2$ and $x^3$. Thus the space-time metric is given by (37) and we get by (41) and (42) :

$$\Gamma'_{000} = \frac{1}{2}\left[\left(\frac{p_e}{p_e^\infty}\right)^2\right]_{,0} = \frac{p_e}{c\left(p_e^\infty\right)^2}\frac{\partial p_e}{\partial t}, \qquad \Gamma_{i00} = -\frac{1}{2}\left[\left(\frac{p_e}{p_e^\infty}\right)^2\right]_{,i} = -\frac{p_e}{\left(p_e^\infty\right)^2}\,p_{e,i}\,,$$

$$\Gamma^0_{00} = \left(\frac{p_e}{p_e^\infty}\right)^2\Gamma'_{000} = \frac{1}{cp_e}\frac{\partial p_e}{\partial t}, \qquad \Gamma^i_{00} = -\left(\frac{p_e}{p_e^\infty}\right)^2\Gamma_{i00}. \qquad (63)$$

This implies that Eq. $(62)_2$ is equivalent to

$$p_{e,\,i} = 0 \quad (i = 1,...,3)\,, \qquad\qquad (64)$$

which means that the field $p_e$ has a uniform value $p_e^0$ in the massive body, occupying a domain $\Omega$ in the space M. This result demands that we are able to give a sense to our basic assumption (A) in a case where the direction of the field **g** , asked for by Eq. (16), is undetermined. We restrict the discussion of this degenerate situation to the case where the field $p_e$ is furthermore uniform at infinity. Then, either the uniform value $p_e^0$ is $p_e^\infty$, in which case there is no space contraction in the considered domain, which must extend to infinity. Or $p_e^0 < p_e^\infty$, in which case the domain $\Omega$, where $p_e \equiv p_e^0$, has to be bounded. Thus, the direction of contraction is determined at the boundary $\partial\Omega$, by continuity with the exterior of $\Omega$ where grad $p_e$ is not nil, and the contraction ratio $\beta = p_e/p_e^\infty$ has the same value at each point of $\Omega$ and $\partial\Omega$. This may be sufficient to determine a unique direction of contraction at any point of $\Omega$ It is at least true in the present case of spherical symmetry, where the direction must obviously be the radial one. Now let us examine whether the condition $(62)_1$ gives a further requirement which should be satisfied by dust matter comoving with ether, in addition to (64). With Eqs. (63) and (64), we calculate

$$\frac{dx^0}{ds} = \frac{p_e^\infty}{p_e}\,, \quad \frac{d^2x^0}{ds^2} = \frac{p_e^\infty}{p_e}\frac{d}{dx^0}\left(\frac{p_e^\infty}{p_e}\right) = -\frac{(p_e^\infty)^2}{p_e^3}\frac{dp_e}{dx^0} \qquad (65)$$
and

$$\Gamma'^0_{00}\left(\frac{dx^0}{ds}\right)^2 = \frac{1}{p_e}\frac{dp_e}{dx^0}\left(\frac{p_e^\infty}{p_e}\right)^2 \,. \qquad (66)$$

Hence $(62)_1$ is automatically verified if $p_e$ depends only on t within the body.

Let the free fall begin at time t=0 : the initial distribution $p_{e0}$ must be independent of r. This is not compatible with a static situation at t=0 since, due to Eq. $(55)_1$ (with C=0), it would mean that the Newtonian potential $U_N(r)$ does not depend on r inside the body - that is, for r<R - and this in turn implies that $\rho_0(r)$ is nil for all r. *The collapse in free fall cannot start from a static situation.* The space-time metric $\boldsymbol{\gamma}$ is given by Eq. (37) with $a_1^0 = 1, a_2^0 = r^2, a_3^0 = r^2 \sin^2 \theta$, in particular $\gamma_{0i} = 0$ (this is general in the frame $E$, Sect. 4). For dust matter, the mixed components of the energy-momentum tensor in the comoving frame $E$ are therefore

$$T_\lambda^\mu = \delta_{\lambda 0} \; \delta_{\mu 0} \; u^0 \; u_0 \; \rho \; c^2 = \delta_{\lambda 0} \; \delta_{\mu 0} \; \rho \; c^2 \; , \quad (67)$$

and $T^{\lambda\mu} = T_\lambda^\mu / \beta^2$ with $\beta = p_e(t)/p_e^\infty$ for the contravariant ones; it follows that only the component $\alpha=0$ of the CE for **T** is not trivially verified, and gives

$$\frac{\partial \rho}{\partial t} - \frac{1}{\beta} \frac{d\beta}{dt} \rho = 0 \; , \; \text{i.e. } \rho = f(r) \, \beta(t) \; . \quad (68)$$

We thus get :

$$\rho(t, r) = \frac{p_e(t)}{p_e^\infty} \frac{p_e^\infty}{p_e(t=0)} \rho(t=0, r) = \frac{p_e(t)}{p_{e0}} \rho_0(r) \; . \quad (69)$$

Inserting (69) into the field equation (54) in which $p_e$ depends only on t, we obtain :

$$\frac{d}{dt}\left( \frac{1}{p_e} \frac{dp_e}{dt} \right) = -a \, p_e^{\; 3} \; , \quad a = \frac{4\pi \, G \rho_0}{\left( p_e^{\; \infty} \right)^2 p_{e0}} \; , \quad (70)$$

or after substitution of the proper time $\tau$, which flows uniformly in the body ($d\tau/dt = p_e/p_e^\infty$):

$$\frac{d^{\, 2} p_e}{d\tau^{\, 2}} = -b \, p_e^{\; 2} \; , \quad b = \frac{4\pi \, G \rho_0}{p_{e0}} \; . \quad (71)$$

Note that the constancy of $a$ and $b$, i.e. the fact that the initial mass-energy density $\rho_0$ actually does not depend on r, is a consequence of the field equation with a uniform field $p_e(t)$ inside the body, but in turn, by (74), implies that $\rho$ remains uniform : *the collapse in free fall can occur only with a uniform material density $\rho$*. We thus have already two important differences with GR.

The resolution of Eq. (71) is standard : one introduces the slope $z=dp_e/d\tau = z(p_e)$ as an auxiliary function, considered as one of $p_e$, and one writes

$$\frac{d^2 p_e}{d\tau^2} = \frac{dz}{d\tau} = z \frac{dz}{dp_e} \quad , \qquad (72)$$

whence with (71) (where $b=1$ by changing the time unit) :

$$z \, dz + p_e^2 \, dp_e = 0 \ , \quad z^2 + (2/3) \, p_e^3 = C, \quad (73a)$$

$$z = \frac{dp_e}{d\tau} = \frac{+}{-} \ \sqrt{C - \frac{2}{3} p_e^3} \quad , \qquad\qquad (73b)$$

which gives $\tau$ as function of $p_e$ :

$$\tau(p_e) = D \ \frac{+}{-} \int_0^{p_e} \frac{dq}{\sqrt{C - \frac{2}{3} q^3}} = D \ \frac{+}{-} \ F(p_e) \quad . \quad (74)$$

One starts with a positive value $p_{e0}$ at $\tau = t = 0$. The initial slope $z_0$ determines the integration constant C, which is positive (Eq. (73a)). In that case, if one restricts the discussion to curves which cannot go accross the ($p_e=0$)-axis, the solution $p_e$ to Eq. (71) is a periodic function of $\tau$ , which cancels at regular intervals $\delta\tau = 2 \ F(p_{e \, max})$ with $p_{e \, max} = (3C/2)^{1/3}$ (Eq. (73a)); at those times $\tau$ where $p_e$ cancels, the slope z undergoes a discontinuity, passing from $-\sqrt{C}$ to $\sqrt{C}$ (Eq. (73b)). Whatever the sign of $z_0 = (dp_e/d\tau)(\tau=0)$, the macroscopic ether pressure $p_e$ inside the body cancels after a finite proper time. Actually, the sign of $z_0$ is related to the time evolution of the radius R' of the body, as evaluated with the measuring rods of the freely falling "observer" (a robust one) : assumption (A) means that R' is expressed as function of the constant radius R (as evaluated with the Euclidean metric bound with ether, as long as this remains comoving with matter) as R' = ($p_e^\infty / p_e(\tau)$) R , hence

$$\frac{dR'}{d\tau} = - \frac{z}{p_e^2} \, p_e^\infty \, R \quad . \quad (75)$$

Thus, if the local "observer" finds an implosion at $\tau=0$, one must have $z_0 > 0$, but then the slope z cancels at the proper time

$$\tau_1 = \int_{p_{e0}}^{p_{e \, max}} \frac{d\tau}{dp_e} \, dp_e = F(p_{e \, max}) - F(p_{e0}) \ , \quad (76)$$

so that the sign of the time evolution of R' is changed : *the implosion becomes an explosion* ! moreover, as $\tau$ continues to increase, $p_e$ cancels at $\tau_2 = \tau_1 + F(p_{e \, max})$ and R'$\to\infty$ as $\tau \to \tau_2$ : the nil value of $p_e = \rho_e c^2$ means that the finite amount of ether contained in the domain r<R, and which is comoving with the body, is diluted in a domain of infinite size - as evaluated with physical rods. Once the value $p_{e\,max}$ has been passed beyond, the density of matter decreases as $p_e$ , since $\rho=\rho_0\,p_e/p_{e0}$ (Eq. (69)). While $\rho$ continues to decrease, say when it becomes much smaller than its initial value $\rho_0$, the argument of Sect. 5, that all ether would be involved in material particles, cannot be used any more to admit that the ether is comoving with matter (but, on the other hand, since all is uniform within the body, any reason to change this situation should have to come from an interaction with the exterior). Thus, the calculations have only an indicative value after $p_e$ and $\rho$ have gone below the values they took at the start of the free fall. *During the raid in the domain of very high densities, however, the comoving of ether and matter is nearly implied by the assumption of the constitutive ether (see Sect. 5) and in turn it precludes that the collapse in free fall lead to any singularity.*

The measuring rods of sufficiently far observers are not affected by gravitation ($p_e \approx p_e^\infty$), thus one could be tempted to say that, as long as its density is high enough so that ether is comoving with matter, the body has a constant radius for remote observers : if the macroscopic motions of ether and matter coincide, a change in the physical size of a massive object can occur only with a change in the gravitational contraction. But as soon as the metric is influenced by gravity - and thus by such objects - it is difficult for a remote observer to define the size of an object, e.g. because the light paths become curved. By the way, can light rays escape the freely falling body in finite time t, or does it become a "black hole" ? A first point is that the time t (of far clocks) becomes *infinite* during the proper time $\tau_2$ (of freely falling clocks in the body or at its surface) that it takes for the value $p_e$ inside the body to cancel, since

$$t(\tau_2) - t(\tau_1) = \int_{\tau_1}^{\tau_2} \frac{dt}{d\tau}\,d\tau = \int_{\tau_1}^{\tau_2} \frac{p_e^\infty}{p_e(\tau)}\,d\tau = \int_{p_{e\,max}}^{0} \frac{p_e^\infty}{p_e}\,\frac{d\tau}{dp_e}\,dp_e = \int_{0}^{p_{e\,max}} \frac{p_e^\infty\,dp_e}{p_e\,\sqrt{C - 2\,p_e^3\,/\,3}} = +\infty. \ (77)$$

However, to answer the question about light rays, we should be able to solve the equation

$$\frac{dt}{dr} = \frac{dt}{dl^0} = \frac{dt}{d\tau}\frac{d\tau}{dl}\frac{dl}{dl^0} = \frac{1}{c\,[\beta(t,\,r)]^2} \ , \ \ t(r{=}R) = t_0 \geq 0 \ \ \ (\beta = \frac{p_e}{p_e^\infty}) \ , \ \ (78)$$

which implies that we should know the field $p_e(t, r)$. In GR, Birkhoff's theorem allows one to choose space-time coordinates in which the exterior metric is static, and thus to find ones in which it is the Schwarzschild metric. This means that, according to GR, gravitation waves cannot exist outside the massive body, in the case of spherical symmetry (see e.g. Weinberg [4]). *But it is not so in the proposed theory*. The field equation (54) is a hyperbolic quasi-linear differential equation in the variables (t, r), since it writes

$$A \frac{\partial^2 f}{\partial t^2} + 2B \frac{\partial^2 f}{\partial t \, \partial r} + C \frac{\partial^2 f}{\partial r^2} + D = 0, \;\; f = \beta^2 = \left( \frac{p_e}{p_e^{\infty}} \right)^2,$$

$$A = \frac{-1}{c^2 f}, B = 0, C = f, D = 2 \frac{f}{r} \frac{\partial f}{\partial r} + \frac{1}{c^2 f^2} \left( \frac{\partial f}{\partial t} \right)^2 - \frac{8\pi G}{c^2} \rho f \;. \quad (79)$$

The characteristics are therefore the radial light rays :

$$\frac{dr}{dt} = \frac{+}{-} c \left[ \beta(t, r) \right]^2, \quad (80)$$

(which shows in a demonstrative way that gravity propagates with the velocity of light in the ETG, though it is already clear from the general equation (23)). Thus, depending on the initial and boundary conditions, the solution to (79) is generally made of both outgoing and ingoing waves. In the present case, the initial condition $f(t=0, r) = f_0(r)$ and $(\partial f / \partial t)(t=0, r) = q_0(r)$ for all $r \geq 0$ determines uniquely the solution in the domain $r \leq \phi(t)$, where $(r = \phi(t))$ is the outgoing characteristic starting from the origin $(t=r=0)$ [20], and we have seen that the initial data cannot be static, that is, $q_0(r) \neq 0$. Therefore, a non-static situation (with waves) is certainly present for $t > 0$, and since it resolves in a uniform (but time-dependent) situation for $r < R$, we may confidently expect that outgoing *and* ingoing waves are present. Now we observe that the wave velocity $w = |dr/dt| = cf = c\gamma_{00}$ *decreases* along the path of the ingoing wave, because the wave moves towards the direction where $p_e$ decreases and the clocks are slowed down; we have in general :

$$\frac{dw}{dr} = c \frac{\partial f}{\partial r} + \frac{\varepsilon}{f} \frac{\partial f}{\partial t} \;\; \text{on} \;\; (\frac{dr}{dt} = \varepsilon c f), \; (\varepsilon = \pm 1), (81)$$

thus our statement is certainly true inside the body ($r \leq R$) for $\tau \geq \tau_1$ (Eq. (76)), since we have then $\partial f / \partial r = 0$ and $\partial f / \partial t < 0$, whence $dw/dr > 0$ for the ingoing wave ($dr < 0$, $\varepsilon = -1$); this strict inequality extends to a larger domain $r < R + h(t)$; the inequality is also ensured at sufficiently large $r$, where the wave starts from a static situation. But a wave whose velocity decreases along its direction of propagation may become a shock wave if its propagation length is sufficient, as discussed by Whitham [20] (the other condition ensuring the appearance of a shock wave is that the decrease in the wave speed should be stronger than the damping of the wave; intuitively, we would say that here an ingoing wave should be reinforced rather than damped, but this must be checked). If we admit on the above basis that $w$ decreases along ingoing characteristics in the whole domain $r \leq \phi(t)$ (and that ether is still comoving with matter), then we have necessarily enough time $t$ for a shock wave to develop before the "catastrophe" $p_e = 0$, since it takes an infinite time $t$ [Eq. (77)] [the catastrophe relies precisely in the fact that a finite proper time corresponds to an infinite time of remote clocks (and the reverse for the spatial size), a very strange situation]. According to our theory based on a constitutive ether, a shock wave, i.e. a discontinuity in the ether pressure, should destroy the material particles made of stable local flows in ether (perhaps vortices). Thus, once a shock wave would have reached the body, an enormous amount of energy would be released, thereby ending the collapse (otherwise, the whole mass would be destroyed) and at the same time reducing the mass of the body, i.e. the cause of the collapse. Winterberg already suggested the possibility of a shock wave in gravitational collapse, within his different theory [7] (see below). This implosion-explosion cycle could repeat until the mass is small enough to allow static equilibrium. These considerations might be of interest for the interpretation of $\gamma$ bursts.

## 9. CONCLUSION

Starting from a tentative interpretation of Newtonian gravity as a pressure force in an incompressible perfect fluid or ether, one may first obtain a new theory of gravity within Newtonian mechanics, simply by giving a compressibility to the ether; this results in a finite velocity of propagation. This does not seem to have been proposed earlier, and the reason may be that one : surprisingly, the "ether pressure" $p_e$ must decrease towards the gravitational attraction whereas the average mass density in the particles depends on $p_e$ only and decreases with $p_e$, and thus also decreases towards the attraction - thus the particles would swell in the gravitation field, i.e. there, where they are closer from one another! But the density in the particles is not the density of matter (the number of particles per unit volume), hence there is no absurdity and moreover one sees then at a glance that matter cannot be indefinitely compressed (since the particles would then encroach up on one another). When one accounts for the metric effects of gravity, the essential point is the coexistence of two spatial metrics in the reference frame $E$ of ether : an Euclidean one $\mathbf{g}^0$, which is bound to the ether (when the latter is considered at a macroscopic scale), i.e. which makes it a rigid "body"; and the physical one $\mathbf{g}$, which is related to $\mathbf{g}^0$ by a contraction of measuring rods in the ratio $p_e/p_e^\infty$ (with $p_e^\infty$ the value of $p_e$ in remote regions, free from gravitation), depending on time in the non-static situation. The clocks bound with $E$ are assumed to be slowed down in the same ratio, and special relativity allows to pass to moving local frames. This assumption of gravitational contraction (resp. dilation) of space (resp. time) standards expresses the principle of equivalence between the metric effects of motion and gravitation, which has a basis in the ETG. Independently of any field equation, it allows, *in the static case*, to deduce the geodesic characterization of motion of ordinary as well as light-like particles from Newton's second law expressed with physical standards affected by gravitation.

The field equation for $p_e$ obtained within Newtonian mechanics in assuming an ether compressibility keeps the same form in the modification integrating the metric effects, but it becomes non-linear in $p_e$. In the spherical static situation, it leads to Schwarzschild's space-time metric $\gamma$ outside the body, but the interior metric is different and agrees with the EP. Furthermore, the associated "acceleration" field $\mathbf{g}$ (Eqs. (5) and (20)) is the exact Newtonian one, inside and outside. The same interior metric as here has been found (in the particular case of uniform density) by Winterberg in his theory, also based on ether [7]. However, Winterberg's theory is very different, first in the concept of ether : he assumes that the ether is made of quantum particles, obeys (nonrelativistic) mechanics and is subjected to gravitation. Here it is continuous at any scale, the macroscopic ether of the gravitation theory defines a global inertial frame (when there is one), and gravity is caused by ether as its pressure force. Winterberg's theory needs ether plus an independent absolute space which would be a Newtonian inertial frame for the moving ether (Winterberg's ether has a macroscopic motion even in the static case). Winterberg's field equation also is very different, since it is one for a 4-vector potential (instead of scalar here) and it is linear in this potential, very much like the equations for the electromagnetic vector and scalar potentials.

The basic assumption that the material particles are local flows in the fluid ether [13] leads to admit that, in the case of very high density, the macroscopic motions of ether and matter coincide. The study of gravitational collapse in free fall (CFF) with spherical symmetry, made under this assumption, exhibits marked differences with GR. If nothing would stop the free "fall", the implosion would necessarily be followed by an explosion in such a way that, as seen from local observers, the space within the body would become infinite. This is the exact contrary of the point singularity of GR; perhaps our calculation is less justifiable when the density ceases to be very high, but it does apply to preclude the occurrence of a singularity with infinite density. Moreover, according to the proposed theory, gravitation waves (moving at the velocity of light) may well exist outside a body also in the case of spherical symmetry, and indeed in the situation of CFF the occurrence of an ingoing shock wave seems quite plausible, which would release energy and stop the free fall.

It is emphasized that the present theory is non-covariant; this is logically consistent since it assumes a privileged frame. But therefore, the fact that Schwarzschild's exterior metric is found in the static case with spherical symmetry does not completely assure that such observational data as the advance in the perihelion of planetary orbits are satisfactorily predicted : it is necessary to build an equivalent of the post-Newtonian (pN) approximation of GR and to check that it agrees with observational data. Progress is currently being made in this direction and the results obtained so far seem to remain compatible with this requirement, although a further investigation is still needed. In particular, this research in progress has already shown that no gravitation wave can be found in the first pN correction. The pN approximation also shows that the theory with Newtonian space and time (Sect. 2) is not a consistent approximation of the present theory, since the effects of the compressibility cannot be separated from the metric effects.

A recent result is that a consistent Newton law can also be defined in the general (time-dependent) situation, and leads to a true conservation equation for the energy; this includes both the (negative) potential energy of matter in the gravitation field and the positive energy of the gravitation field itself. However, the Newton law is *not* compatible with geodesic motion in time-dependent situations. The energy conservation equation, together with the weak-field, linear approximation of the present theory, allow to calculate the gravitational radiation flux in a way similar to that followed in linearized GR. In particular, if the center of mass of the gravitating system is at rest in the privileged frame or "ether", this radiation flux corresponds to an energy *loss* which depends only on the third time derivative of the quadrupole tensor and involves a $G/c^5$ factor (this is much as the famous "quadrupole formulae" of GR, on which are based the estimates of the secular decrease in the orbital period for a binary system). For this question also, however, it remains to verify that the effect of changing the reference frame is small for expectable absolute velocities.

Let us finally comment on the Lense-Thirring effect: this is a prediction of GR, according to which the rotation of a massive body should manifest itself by a "magnetic" component of the gravitation field, thereby causing a specific precession of the orbit of any test particle around the body. This effect comes from the fact that in GR: (i) particles move along geodesics and (ii) non-zero $\gamma_{0i}$ components of metric $\boldsymbol{\gamma}$ are predicted in the case of stationary rotation [4,18]. In the present theory, a stationary rotation of a massive body, with respect to the privileged frame $E$, causes a constant field of "ether pressure" $p_e$ and, in that case, this theory also predicts geodesic motion of test particles- but the $\gamma_{0i}$ components are always zero in the frame $E$. This frame is inertial in the sense that the Newton law is written in $E$, and for constant field $p_e$ it is rigid with respect to the physical space metric, in short it is an inertial frame in the astronomical sense. Hence, the Lense-Thirring effect *does not exist according to the present theory*. But the frame $E$ is defined by the average motion of the assumed microscopic *constitutive* ether. Therefore, the rotation of the massive body, as any motion of matter, contributes to the motion of the frame $E$, by a simple weighting effect. This means that the theory is in agreement with Mach's principle, because the preferred inertial frame (playing the role of Newton's absolute space, and indeed exactly the same role in the limit of weak and slowly varying fields) is influenced by all motions of material bodies.

ACKNOWLEDGEMENTS

I am very grateful to Prof. P. Guélin for many helpful discussions and comments on this work. It is also a pleasure to thank Prof. E. Soos for his interesting remarks on this subject. The interest of the questions of one referee, on the gravitational radiation and the Lense-Thirring effect, is also acknowledged.